\documentclass[11pt]{revtex4-1}

\usepackage{graphicx}

\usepackage[latin1]{inputenc}
\usepackage{amsmath}
\usepackage{amsfonts}
\usepackage{amssymb}
\usepackage{makeidx}
\usepackage{color}
\usepackage{mathrsfs}
\usepackage{amsfonts}
\RequirePackage[colorlinks,citecolor=blue,urlcolor=magenta,linkcolor=blue]{hyperref}

\begin{document}

\title{Formation Of Emergent Universe in Brane Scenario as a Consequence of Particle Creation}

\author{Jibitesh Dutta\footnote {jdutta29@gmail.com,jibitesh@nehu.ac.in}}
\affiliation{Department of Basic Sciences, North Eastern Hill University, Shillong 793022, Meghalaya, India.}

\author{Sourav Haldar\footnote {sourav.math.ju@gmail.com}}

\affiliation{Department of Mathematics, Jadavpur University, Kolkata 700032, West Bengal, India.}

\author{Subenoy Chakraborty\footnote {schakraborty@math.jdvu.ac.in}}

\affiliation{Department of Mathematics, Jadavpur University, Kolkata 700032, West Bengal, India.}


\begin{abstract}
Here we formulate scenario of emergent universe from particle
creation mechanism in  spatially flat braneworld models. We
consider an isotropic and homogeneous universe in Braneworld
cosmology  and  universe is considered as a non-equilibrium
thermodynamical system with dissipation due to particle creation
mechanism. Assuming the particle creation rate as a function of
the Hubble parameter , we formulate emergent scenario in RS2 and
DGP models of Braneworld.

\end{abstract}
\maketitle
Keywords: Emergent Universe, Modified Gravity, Braneworld, Particle Creation, Non-equilibrium Thermodynamics\\\\
PACS Numbers: 98.80.Cq, 98.80.-k


\section{Introduction}
The existence of big bang singularity in the very early universe
remains an open issue although  most of the mysteries of standard
hot
 big bang cosmological  model have been addressed by inflationary epoch
 \cite{Guth:1980zm}.
To overcome this initial singularity, Ellis $et$ $al$ proposed  a
scenario called Emergent Universe, which is ever existing without
singularity and has an almost static behaviour in the infinite
past \cite{Ellis:2002we,Ellis:2003qz}. Eventually the model
evolves into an inflationary phase.

Recently there have been  lot of interest in emergent universe
models based on  standard as well as modified gravity
\cite{Zhang:2013ykz, Bag:2014tta,Chakraborty:2014ora,
Paul:2014yga} . This resurgence of interest
 is because of CMB observations favouring early inflationary universe \cite{Ade:2013uln,Labrana:2013oca}
   and probably this is the possible mechanism for present acceleration \cite{sp}.
  Historically the credit goes
  to Harrison \cite{Harrison:1967zz}
 for obtaining  a model of closed universe  with radiation  and he showed  that  asymptotically it approached  Einstein static universe
  (ESU).  Then after a  long gap Ellis and Maartens \cite{Ellis:2002we,Ellis:2003qz} in recent past were able to formulate closed universe
  with a minimally coupled scalar field $\phi$.
   However, exact analytic solutions were not presented in their work and only asymptotic behaviour  agreed with emergent universe.
   Subsequently, Mukherjee $et$ $al$ \cite{Mukherjeea} obtained solution for
   Starobinsky  model with features of an emergent universe. Also Mukherjee $et$ $al$ \cite{Mukherjee:2006ds} formulated a general framework for
    an emergent universe
   using an adhoc equation of state which has exotic behavior in some
   cases. Very recently the idea of quantum tunnelling has been
   used to model emergent universe \cite{Labrana:2011np}.

            In most of the General Relativity (GR) based models, Emergent universe is generally
           obtained for spatially closed past ESU which
                     is not stable . Contrary to this in modified gravity, emergent
           scenario can be obtained in spatially flat as well as
           open/closed model of universe \cite{Zhang:2013ykz, Bag:2014tta}.
          Also in the context of modified gravity, including Braneworld
           models  it has been found, a stable ESU can be obtained \cite{Gruppuso:2004db}. It may be noted that
           a stable ESU also exists in an open universe in
           $f(T)$ gravity  , loop quantum  cosmology and Horova Lifshitz gravity
           \cite{Canonico:2010fd,Wu:2011xa,Parisi:2012cg}. Since
            astronomical observations including very recent Planck  results
            favour spatially flat universe it is imperative to
            examine Emergent scenario in spatially flat universe. In this paper, we look for a possibility  of emergent
            scenario from particle creation mechanism in  spatially flat braneworld models.
            The modern era of brane
cosmology began in the context of large extra dimensions, made
possible by the hypothesis that the standard model of particle
physics is localized on a D-brane.  Braneworld models  of the
universe have been  the focus of  attention  in the current decade
 for the search of many outstanding problems in  Cosmology.  These
 kind of models are inspired by String Theory.
           In braneworld scenario, our
four dimensional universe (a brane) is a hypersurface embedded in
higher dimensional bulk spacetime. In this brane-bulk scenario,
all matter and gauge interactions (described by open strings) are
localised on a brane while gravity (described by closed strings)
may propagate into whole space time. This means that gravity is
fundamentally a higher dimensional interaction and we only see the
effective 4D theory on brane. Among the different proposals for
brane models, the two
 prominent are RS (proposed by Randall and Sundrum) and DGP(proposed by
 Dvali,Gabadaze and Porrati) models. In RS models, the hierarchy
 problem could be solved by a warped or curved extra dimension
 showing that fundamental scale could be brought down from the
 Planck scale to 100 GeV.

While $RS2$ model have only {\em one} $(1 + 3)$-brane \cite{rs2},
the  $RS1$ model have two  {\em two} $(1 + 3)$-branes at the ends
of the orbifold $S^1/Z_2$ \cite{rs1}. Due to its simple and rich
conceptual base, RS2 model  is very popular and  got much
attention \cite{pb00, pbd00, cg, cgs} and  in particular, on the
inflationary scenario.

The main idea of the DGP model is the inclusion of a
 four dimensional Ricci-Scalar into the action .On the
4-dimensional brane the action of gravity is proportional to
$M_{P}^{2}$ whereas in the bulk it is proportional to the
corresponding quantity in 5-dimensions. The model is then
characterized by
 a cross over length scale
     $$ r_{c}=\frac{M_{P}^{2}}{2M_{5}^{2}}$$
such that gravity is 4-dimensional theory at scales $ a\ll r_{c}$
where matter behaves as pressure less dust but gravity
\textit{leaks out} into the bulk at scales $ a\gg r_{c} $ and
matter approaches the behaviour of a cosmological constant
\cite{dgp1,dgp2,dgp3}.

 In the standard cosmology RS2 model modifies the early universe
 where as DGP gravity modifies the late universe. So in the present
 universe brane corrections are not effective in RS2 model. But
 as phantom energy increases with the expansion, it brings drastic
 changes in the RS2 model  \cite{sksjd}.  This interesting and recent
 feature of RS2  model which can modify late time cosmic expansion
(if  the energy density of the matter content increases at late
time  e.g., phantom scalar field) has also been  beautifully shown
by using dynamical system tools in ref
\cite{GarciaSalcedo:2010wa}.
 But this kind of modification does
 not appear  if the energy density of the matter content decreases with cosmic expansion e.g, quintessence scalar field, radiation, dust etc.
  Thus RS brane model can also modify late time cosmic expansion in addition to its appreciable impact on early universe cosmology.

  Very recently Chakraborty in ref \cite{Chakraborty:2014ora} formulated a emergent universe
  model in   GR using the mechanism of  particle creation. The aim of this paper is to formulate an Emergent Scenario
  in Braneworld
  as consequence a of particle mechanism. In a sense this paper is
  a   generalisation of the one reported in \cite{Chakraborty:2014ora} to include higher
  dimensional behaviour. Often  using BG  one obtains cosmological surprises.
    It may be noted that  from thermodynamical aspect it
   has been proposed that entropy consideration favors the
   Einstein static phase as the initial state of our universe
   \cite{Gibbons:1987jt,Gibbons:1988bm}.
   The paper is organized as follows :
      Section 2 deals with  non-equilibrium thermodynamics from the perspective of particle creation in cosmology
      while in section 3
       we present Emergent scenario in Braneworld models.
      The paper ends with a short discussion and Concluding Remarks 4.

\section{Non Equilibrium Thermodynamics and Particle Creation in cosmology}
Suppose $E$ be the internal energy of a closed thermodynamical
system having $N$ particles. The first law of thermodynamics which
is essentially the conservation of internal energy is given by
\cite{Prigogine:1989zz}
\begin{equation}
dE=dQ-pdV
\end{equation}
\noindent where as usual $p$ is thermodynamic pressure, $V$ any
co-moving volume and $dQ$ represents the heat  received  by the
system in time $dt$.

From the  above conservation equation, we get the Gibb's equation
given by

\begin{equation}\label{eq2}
T ds = dq =d\Big(\frac{\rho}{n}\Big)+p~d\Big(\frac{1}{n}\Big)
\end{equation}

\noindent $n=\frac{N}{V}$ , $dq= \frac{dQ}{N}$ and
$\rho=\dfrac{E}{V}$ being the particle number density , heat per
particle and energy density of the system respectively. It may be
noted that this equation is also true when the particle number is
not conserved i.e., the system is not a closed system
\cite{Harko:2012za}.
 Here we shall consider an open thermodynamical system where the number of fluid  particles  is not conserved
 and the non conservation  of fluid particles  is expressed as ($N_{;\mu}^{\mu}\neq 0$)

 \begin{equation}\label{eqthree}
 \dot{n}+ \Theta n= n\Gamma
\end{equation}
where $N^{\mu}=n u^{\mu}$ is the particle  flow vector, $u^{\mu} $
is the particle four velocity, $\Theta = u_{;\mu}^{\mu}$ is the
fluid expansion, $\Gamma$ stands for  the rate of change of the
number of particles  ($N=na^3$) in a co-moving  volume  $V=a^3$
and by notation $ \dot{n}=n_{,\mu} u^{\mu}$.

The positivity of $\Gamma$ indicates creation of particles  while
there is  annihilation  of particles  for negative $\Gamma$ . Any
non zero $\Gamma$  will behave as  an effective bulk pressure  of
thermodynamical fluid and one can use non equilibrium
thermodynamics \cite{Chakraborty:2014pca}.

In this work, we shall consider a flat, homogeneous and isotropic
model of the Universe, given by the
Friedmann-Lemaitre-Robertson-Walker (FLRW) metric
 in comoving coordinates ($t$, $r$, $\theta$, $\phi$)  as

 \begin{equation}\label{eq4}
    ds^2 = dt^2-a^2(t) \left[dr^2+r^2(d\theta ^2+\sin^2\theta d\phi
^2)\right],
\end{equation}
\noindent where $a(t)$ is the scale factor of the Universe. Here
we consider universe as an  open thermodynamical system which is
non equilibrium in nature due
 to particle creation.\\\\

\noindent The energy momentum tensor  of the cosmic fluid  is
characterized  by the energy momentum tensor
\begin{equation}\label{eq5}
T_{\mu\nu} =(\rho +p+\Pi) u_\mu u_\nu +(p+\Pi)g_{\mu\nu}
\end{equation}

\noindent The  energy conservation relation of  $ T^{\mu \nu}_{;
\nu}=0$ takes the following form

\begin{equation}\label{eqsix}
\dot \rho+3H(\rho+p+\Pi)=0
\end{equation}
\noindent where $ H=\dfrac{\dot a}{a}$ is the Hubble parameter.
Here the pressure term $\Pi$ is related  to some dissipative
phenomena (say bulk viscosity). In the present context, however
the cosmic fluid may be considered  as perfect fluid where
dissipative term $\Pi$ is the effective bulk pressure due to
particle creation or equivalently  the conventional dissipative
fluid  is not taken as cosmic substratum , rather  a perfect fluid
with  varying particle number  is considered. For an isentropic
(or adiabatic ) particle production  this equivalence can  be
nicely described as follows
\cite{Chakraborty:2014pca,Zimdahl:1996ka}:

Using the conservation eqs. ~(\ref{eqthree}) and (\ref{eqsix}) the
entropy variation can be obtained  from Gibb's  equation (2) as

\begin{equation}\label{eq7}
    n T \dot{s}= -3H\Pi- \Gamma(\rho+p)
\end{equation}

\noindent where  $T$ is the temperature of the fluid.

 But due to isentropic   (or adiabatic ) process, the equilibrium
 entropy per particle does not change (as it does in dissipative
 process), i.e, $\dot{s}=0$, and from eq.(\ref{eq7}) the effective bulk pressure  is
 determined  by particle creation  rate as
 \begin{equation}\label{eq8}
\Pi=-\frac{\Gamma}{3H}(\rho+p)
\end{equation}

\noindent Hence, the bulk viscous pressure is entirely determined
by particle production rate. So we may say that a dissipative
fluid is equivalent to a perfect fluid having varying  particle
number. It may be noted that, although $\dot{s}=0$, still there is
entropy production  due to  enlargement  of the phase space of the
system. In the present context the same effect is due to the
expansion of the universe. This effective bulk pressure does not
correspond to conventional non equilibrium  phase,  rather a state
having  equilibrium properties as well. We note that it is not the
equilibrium era with  $\Gamma=0$

From eq.(\ref{eq8}), we see that if we know  the Hubble parameter
$H$, then the particle creation rate can be  determined. Further
using Friedman equations  one can relate  $\Gamma$ to the
evolution  of universe. In what follows as in
\cite{Chakraborty:2014pca}, assuming the particle creation rate as
a function of  the Hubble parameter , we can study the
cosmological evolution in Braneworld.

\section{ Emergent universe in  Braneworld scenario}

\subsection { RS2  model}
In a homogeneous and isotropic model of universe  given by FRLW
metric (\ref{eq4}), the modified Friedmann equation of RS2 model
is given by

\begin{equation}\label{eq9}
H^2 = \left(\frac{\dot a}{a} \right)^2 = \frac{1}{3}
\rho\left[1+\epsilon\frac{\rho}{2\lambda}\right],
\end{equation}
\noindent where   $\epsilon=\pm 1$ and the sign of $\epsilon$ are
related to positive and negative brane tension respectively.(For
simplicity we are using $8 \pi G =1$ )

In the absence of bulk pressure, the matter content sector satisfy

\begin{equation}\label{eq10}
 \dot \rho+3H(\rho+p)=0,
\end{equation}
In order to highlight some characteristics of this model, we
review  our paper \cite{sksjd} for  the case $\epsilon=-1$ with
$p=\rho \omega$, $\Pi=0$ and $\omega<-1$. Using eqs. (\ref{eq9})
and (\ref{eq10}), we see that exact solution of energy density is
given by

\begin{equation}\label{eq11}
    \rho(t)=\left[-\epsilon\frac{1}{2\lambda} +\left\{\sqrt{\frac{1}{\rho_{0}}+\epsilon\frac{1}{2\lambda}}+\sqrt{\frac{3}{4}}\omega(t-t_{0})
    \right\}^{2}\right]^{-1}
\end{equation}

\noindent where $\rho_0<2\lambda$ is the current energy density of
dark energy and $t_0$ is the present time. From above equation we
see that the braneworld model with $\epsilon=1$ $(\lambda>0)$ that
describes  an accelerated expansion ends in big-rip singularity in
time

\begin{equation}\label{eq12}
    t_s=t_0 -\frac{2}{\sqrt{3}|1 + w|} \Big[\sqrt{\frac{1}{2\lambda}}+\sqrt{\frac{1}{\rho_0} +
  \frac{1}{2\lambda}}\Big]
\end{equation}

\noindent as obtained in most of the GR based models. Brane
gravity corrections make changes in the behaviour  of phantom
fluid  in case phantom fluid dominates RS2 model of the universe
with $\epsilon=-1$ $(\lambda<0)$. In this case it is found  that
it explains the present cosmic acceleration and exhibits
deceleration after a finite time, when energy density grows
sufficiently \cite{sksjd}.

In the present work we take $\Pi \ne 0$, the eq. (\ref{eq9}) can
be rewritten by introduction of  dimensional variable
$x=\rho/2\lambda$

\begin{equation}\label{eq13}
    3H^2= \rho_\textrm{eff}
 \end{equation}

 \noindent where the effective  density is given by
\begin{equation}\label{eq14}
     \rho_\textrm{eff}=2\lambda x(1+\epsilon x)
 \end{equation}
  It is  worth noting that when $x<<1$, the GR based cosmology is
  recovered and brane effects give new kind of behaviour when
  $x>>1$. Acceleration and deceleration of this model without bulk
  pressure have been studied in \cite{sksjd,Srivastava:2007fc} . The
  case of $\epsilon=-1$ is (also known as bouncing universe) have
  been extensively studied in \cite{Shtanov:2002mb,Hovdebo:2003ug}. Also the
  effect of bulk brane viscosity in this model in the framework of
  Eckart's theory has been studied in \cite{Lepe:2008eu}.

  Taking derivative of eq. (\ref{eq13})  w.r.t cosmic time $t$ and using (\ref{eqsix}) , we get

  \begin{equation}\label{eq15}
    \dot{H}=-\frac{1}{2}[\rho_\textrm{eff}
    +p_\textrm{eff}+\Pi_\textrm{eff}]
\end{equation}
where effective pressures are given by
\begin{equation}\label{eq16}
    p_\textrm{eff}=2\lambda[\omega x (1+\epsilon 2x)+\epsilon x^2]
\end{equation}
\begin{equation}\label{eq17}
\Pi_\textrm{eff}=\Pi(1+\epsilon 2x)
\end{equation}
and the effective equation of state is given by

\begin{equation}\label{eq18}
    \omega_\textrm{eff}=\frac{p_\textrm{eff}+\Pi_\textrm{eff}}{\rho_\textrm{eff}}=\frac{1}{1+\epsilon
    x}\left[(1+ 2\epsilon x)\left(\omega+\frac{\Pi}{2\lambda
    x}\right)+\epsilon x\right]
\end{equation}
Adding eqs. (\ref{eq14}) , (\ref{eq16})  and (\ref{eq17})  and
using (\ref{eq8})   , we can rewrite eq (\ref{eq15})  as
\begin{equation}\label{eq19}
2 \dot{H}=-(1+\epsilon 2x)(1+\omega) 2\lambda
x\Big(1-\frac{\Gamma}{3H}\Big)
\end{equation}
Again rewriting modified Friedmann equation (13) in terms of $x$,
we get
\begin{equation}\label{eq20}
2\epsilon\lambda x=\epsilon\lambda[B_{\textrm{RS}}(\epsilon,
H)-1]~~\textrm{where}~~B_{\textrm{RS}}(\epsilon ,
H)=\sqrt{\frac{6\epsilon H^2}{\lambda}+1}
\end{equation}
Substituting the value of $2\lambda x$ in eq. (19), we get
\begin{equation}\label{eq21}
2 \dot{H}=- \frac{6(1+\omega)H^2}{[1+1/B_{\textrm{RS}}(\epsilon,
H)]} \Big(1-\frac{\Gamma}{3H}\Big)
\end{equation}
As the particle creation rate is a function of Hubble parameter
$H$ \cite{Chakraborty:2014pca},  we choose

\begin{equation}\label{eq22}
    \Gamma=3H\Big[{(1-e)+\frac{f}{H}}-\frac{1}{B_{RS}}(e-\frac{f}{H})\Big]
\end{equation}
where $e$ and $f$ are constants.

\begin{figure}
\includegraphics[width=0.5\textwidth]{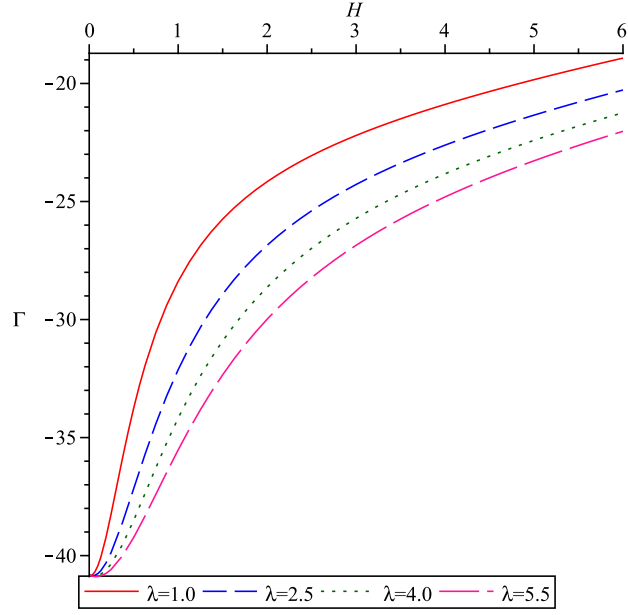}
\caption{Variation of $\Gamma$ with $H$ in RS-II Model,
considering $e=0.787$ and $f=-6.804$.} \label{fig:1}
\end{figure}

Thus using eqs. (\ref{eq21})  and (\ref{eq22}), we get the
differential equation of $H$ describing cosmic evolution as
\begin{equation}\label{eq23}
2 \dot{H}= -6(1+\omega)(eH^2-fH)
\end{equation}

So the Hubble parameter can be obtained by solving eqs. (23), as
 $(\omega\neq-1)$,
\begin{equation}\label{eq24}
\frac{H_0}{H}= H_1+\exp{\Big[-\frac{H_0}{2}(t-t_0)\Big]}
\end{equation}

where $H_0= 6(1+\omega)f~~~,~~~H_1= 6(1+\omega)e$ and $t_0$ is the
 constant of integration. Hence the scale factor can be obtained from
 the above expression for Hubble parameter as
\begin{equation}\label{eq25}
\Big(\frac{a}{a_0}\Big)^{\frac{H_1}{2}}=
{1+H_1\exp\Big[\frac{H_0(t-t_0)}{2}\Big]}
\end{equation}
where $a_0= a(t_0)$

  The above cosmological solution shows the following asymptotic behavior for $f<0$:
    \\$~~~(i) ~a\rightarrow a_0$ , $H\rightarrow 0$ as $t\rightarrow -\infty$
    \\$~~~(ii) ~a\simeq a_0$ , $H\simeq 0$ for $t<<t_0$
    \\$~~~(iii) ~a\simeq \exp[\frac{H_0}{H_1}(t-t_0)]$ , $H\simeq \frac{H_0}{H_1}$ for $t>>t_0$ .

The above asymptotic features show that the above cosmological
solution
 (described by equations (\ref{eq24})  and (\ref{eq25}) ) describes a scenario of emergent
  universe.

\subsection{DGP model}

The modified Friedmann equation in this case is given by
\begin{equation}\label{eq26}
     H^{2}-\epsilon \frac{H}{r_{c}}=\frac{\rho}{3}
\end{equation}
where $ r_{c}=\frac{M_{P}^{2}}{2M_{5}^{2}}$ is the crossover scale
which determines the transition from 4D to 5D behavior and
$\epsilon = \pm 1 $. For $ \epsilon = 1 $, we have standard DGP(+)
model which is self accelerating model without any form of dark
energy, and effective $\omega$ is always non phantom. However for
$ \epsilon = - 1 $, we have DGP(-) model which does not self
accelerate but requires dark energy on the brane. It experiences
5D gravitational modifications to its dynamics which effectively
screen dark energy.

The above equation can be rewritten as
\begin{equation}\label{eq27}
    3H^2= \rho_\textrm{eff} ~~~\textrm{where
    ~effective~density~is} ~~\rho_{\textrm{eff}}= \rho +\epsilon \frac{3H}{r_{c}}
 \end{equation}

Taking derivative of eq. (\ref{eq26})  w.r.t cosmic time $t$ and
using eq. (\ref{eqsix})  , we get

\begin{equation}\label{eq28}
    (2H-\epsilon/r_c)\dot H =-H(\rho+p +\Pi)
\end{equation}

Now using eq. (\ref{eq8})  in above equation we get
\begin{equation}\label{eq29}
    2 \dot{H}=-6 H^2 (1+\omega)\Big(1-\frac{\Gamma}{3H}\Big)B_{\textrm{DGP}}(\epsilon, H)
\end{equation}
where
\begin{equation}\label{eq30}
B_{\textrm{DGP}}(\epsilon,
H)=\frac{H-\epsilon/r_c}{2H-\epsilon/r_c}~~~~~
\textrm{and~it~gives~brane~effects}
\end{equation}

In this case , we choose the particle creation rate as
\begin{equation}\label{eq31}
    \Gamma=
    3H[1-(c-d/H)/B_{\textrm{DGP}}(\epsilon,H)],~~~\textrm{c~and~d~ are~ constants}
\end{equation}

\begin{figure}
\includegraphics[width=0.5\textwidth]{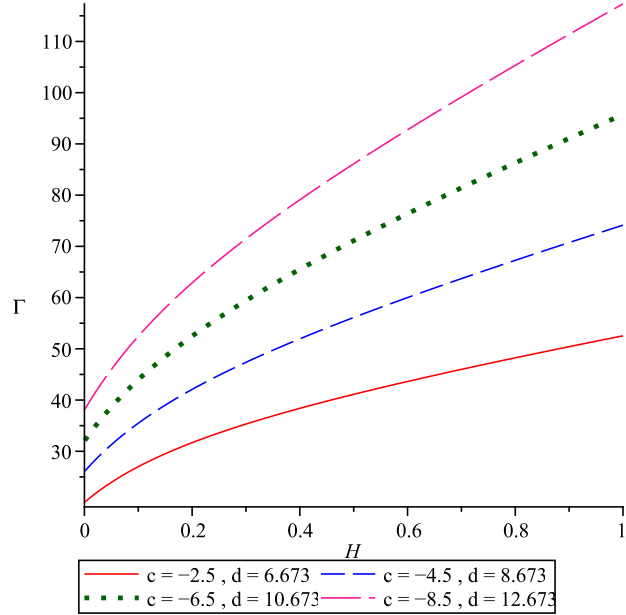}
\caption{ Variation of $\Gamma$ with $H$ in DGP Model, considering
$\epsilon = -1$ and $r_c = 4$ .}\label{fig:2}
\end{figure}

As before using modified Friedmann eqs. (\ref{eq29})  and
(\ref{eq31})  we get the differential equation of $H$ describing
cosmic evolution of DGP model as

\begin{equation}\label{32}
2 \dot{H}=-6(1+\omega)cH^2 +6 d (1+\omega) H
\end{equation}

This evolution equation is identical in form to that for RS-II
brane  gravity (i.e. eq.(\ref{eq23}) ) and hence it is also
poosible to have
 emergent scenario in DGP brane model in the framework of particle
creation mechanism.

\section{Short Discussion and Concluding Remarks}

   The present work deals with spatially flat braneworld models
 in non-equilibrium thermodynamics, based on particle creation
 mechanism. For simplicity, we assume the thermal process to be
 adiabatic in nature and as a result the particle creation rate
 is related to the cosmic evolution. In both RS-II and DGP brane
 model, cosmic fluid with constant equation of state $(p= \omega \rho$ , $\omega \neq -1$)
 is chosen. In both the brane models the particle creation rate
 is chosen as different functions of the Hubble parameter and the
 brane effect is incorporated through the functions $B_{\textrm{RS}}$
 and $B_{\textrm{DGP}}$ respectively. However \, the cosmic evolution
 equation is similar in both the models and we have emergent scenario
 in the early phase of the evolution.
     Further it should be noted that the present model of emergent scenario
 due to particle creation is on ever expanding model of the universe. As the
 particle creation rate is not constant, so the present model has the basic
 difference with the steady state theory of Fred Hoyle $et$ $al$ \cite{Hoyle:1993fh,Hoyle:2000xg} in contrast
 to the emergent scenario in Einstein gravity with constant particle creation
 rate \cite{Chakraborty:2014ora}. Moreover, one can easily check that if we remove the brane effect
 (i.e. $\lambda \rightarrow \infty$ in RS2 or $r_c\rightarrow \infty$ in DGP model)
 then the particle creation rate may be taken to be constant (by choosing $e= 1/2$
 in RS2 model and $c= 1/2$ in DGP brane model). The variation of $\Gamma$ with respect to
 the Hubble parameter has been presented in figs. (\ref{fig:1})   and (\ref{fig:2})  for RS2 and DGP
 brane model respectively for various choices of the parameters involved. Finally,
 we conclude that it is possible to have emergence scenario in brane gravity models
 for non-equilibrium thermodynamics in the framework of particle creation mechanism.

\acknowledgments {The paper is done during a visit to IUCAA, Pune,
India. JD and SC are thankful to IUCAA for warm hospitality and
facility of doing research works.}

  \end{document}